\documentclass[acmsmall,screen]{acmart}
\usepackage{amsmath}
\usepackage{multirow}
\usepackage{booktabs}
\usepackage[table]{xcolor}
\usepackage{array}
\usepackage[table]{xcolor}
\usepackage{graphicx}
\usepackage{adjustbox}
\usepackage{soul}
\usepackage{xcolor}
\usepackage{makecell}
\usepackage{longtable}
\usepackage{footmisc}
\usepackage{hyperref}
\usepackage{pgfplots}
\usepackage{pgfplotstable}
\pgfplotsset{compat=1.18}

\definecolor{lightgreen}{HTML}{C9EAD2}
\definecolor{lightred}{HTML}{F4C7C3}
\usepackage{listings}
\usepackage{xcolor}
\usepackage{caption}

\definecolor{codegreen}{rgb}{0,0.6,0}
\definecolor{codegray}{rgb}{0.5,0.5,0.5}
\definecolor{codepurple}{rgb}{0.58,0,0.82}
\definecolor{backcolour}{rgb}{0.95,0.95,0.92}
\definecolor{diffstart}{rgb}{0.5,0.5,0.5}
\definecolor{diffadd}{rgb}{0,0.5,0}
\definecolor{diffrem}{rgb}{0.6,0,0}

\lstdefinelanguage{diff}{
  morecomment=[f][\color{diffstart}]{@@},
  morecomment=[f][\color{diffrem}]{-},
  morecomment=[f][\color{diffadd}]{+},
  morecomment=[f][\color{diffstart}]{---},
  morecomment=[f][\color{diffstart}]{+++},
}

\lstdefinestyle{DiffStyle}{
    language=diff,
    backgroundcolor=\color{backcolour},
    basicstyle=\ttfamily\fontsize{4.5pt}{4.5pt}\selectfont,
    breaklines=true,
    captionpos=b,
    keepspaces=true,
    numberstyle=\fontsize{4.5pt}{4.5pt}\selectfont\color{codegray},
    numbers=left,
    numbersep=5pt,
    showspaces=false,
    frame=single,
    title={\fontsize}
}

\lstdefinelanguage{json}{
    basicstyle=\ttfamily\fontsize{4.5pt}{4.5pt}\selectfont, 
    numbers=left,
    numberstyle=\fontsize{4.5pt}{4.5pt}\selectfont\color{codegray},
    stepnumber=1,
    numbersep=5pt,
    showstringspaces=false,
    breaklines=true,
    frame=single,
    backgroundcolor=\color{backcolour},
    stringstyle=\color{codepurple},
    literate=
     *{:}{{{\color{diffstart}{:}}}}{1}
      {,}{{{\color{diffstart}{,}}}}{1}
      {\{}{{{\color{black}{\{}}}}{1}
      {\}}{{{\color{black}{\}}}}}{1}
      {[}{{{\color{black}{[}}}}{1}
      {]}{{{\color{black}{]}}}}{1},
}

\pgfplotsset{
    colormap={bluewhitered}{
        color(0cm)=(white);
        color(1cm)=(blue!60!white)
    }
}

\usepackage[framemethod=TikZ]{mdframed}
\definecolor{backgroundpage}{HTML}{F8F5F4}
\newenvironment{summary}[1][]{
    \mdfsetup{
        skipabove=5pt, 
        linecolor=blue,
        roundcorner=2pt,
        innerleftmargin=0.3cm,innerrightmargin=0.4cm, 
        linewidth=0.5pt,
        footnoteinside=false,backgroundcolor=blue!5}
    \begin{mdframed}}
    {\end{mdframed}}

\AtBeginDocument{%
  }

\setcopyright{acmlicensed}
\copyrightyear{2026}
\acmYear{2026}
\acmDOI{XXXXXXX.XXXXXXX}

\begin{document}

\title[Do AI Models Dream of Faster Code?]{Do AI Models Dream of Faster Code? An Empirical Study on LLM-Proposed Performance Improvements in Real-World Software}

\author{Lirong Yi}
\orcid{0009-0008-7536-8367}
\affiliation{%
  \institution{Chalmers University of Technology and University of Gothenburg}
  \city{Gothenburg}
  \country{Sweden}
}
\email{lirongy@chalmers.se}

\author{Gregory Gay}
\orcid{0000-0001-6794-9585}
\affiliation{%
  \institution{Chalmers University of Technology and University of Gothenburg}
  \city{Gothenburg}
  \country{Sweden}
}
\email{greg@greggay.com}

\author{Philipp Leitner}
\orcid{0000-0003-2777-528X}
\affiliation{%
  \institution{Chalmers University of Technology and University of Gothenburg}
  \city{Gothenburg}
  \country{Sweden}
}
\email{philipp.leitner@chalmers.se}

\begin{abstract}
Large Language Models (LLMs) can generate code, but can they generate fast code for complex, real-world software systems? In this study, we investigate this question using a dataset of 65 tasks mined from performance-critical open-source Java projects. Unlike prior studies, which focused on algorithmic puzzles, we conduct experiments on actual performance-sensitive production code and employ developer-written JMH benchmarks to rigorously validate performance gains against human baselines. Our results reveal a nuanced reality---although LLMs demonstrate a surprisingly high capability to solve these complex engineering problems, their solutions suffer from extreme volatility and still lag behind human developers on average. Consequently, we find that the current benchmarks based on algorithmic tasks yields an overly optimistic assessment of LLM capabilities. We trace this real-world performance gap to two primary limitations: first, LLMs struggle to autonomously pinpoint performance hotspots, and second, even with explicit guidance, they often fall short of synthesizing optimal algorithmic improvements. Our results highlight the need to move beyond static code generation towards more complex agent-based systems that are able to profile and observe runtime behavior for performance improvement.

\end{abstract}

\keywords{Large Language Models, Code Generation, Benchmarking, Software Performance, Automated Program Repair}

\received{20 February 2007}
\received[revised]{12 March 2009}
\received[accepted]{5 June 2009}

\maketitle

\section{Introduction}
\label{sec:intro}
Recent advances in generative AI, and specifically Large Language Models (LLMs), have brought the idea of code synthesis into the mainstream of software engineering. Developers now routinely use LLMs for generating requirements~\cite{nouri:24}, test cases~\cite{chen:24}, and refactoring~\cite{liu:25}. We have arguably entered the era of ``vibe coding''~\cite{sarkar:25}, where practitioners generate entire features or applications based on high-level prompts, often focusing on functional correctness while overlooking non-functional implications.

However, one area that still causes concern is the performance efficiency of generated code. While LLMs can generate syntactically correct code, previous work has shown that generated code is often inefficient~\cite{niu:24}, and has noted that standard benchmarks such as CoderEval~\cite{yu:24} lack performance tests to even properly assess whether generated code is efficient~\cite{peng:25}. Further, it has been reported in previous studies that code efficiency, unlike correctness, does not necessarily improve with larger foundation models~\cite{niu:24}. 

Crucially, the current evaluation landscape is insufficient to address this concern. Existing assessments primarily focus on small and isolated algorithmic tasks~\cite{coignion:24, zheng2024beyond}, failing to capture the complexity of performance optimization in real-world software---e.g., managing concurrency models, memory allocations, and cross-file dependencies within massive codebases. To date, it remains an open question whether LLMs can act as effective Performance Engineers in industrial settings. This question is particularly important given that performance is often linked to energy consumption~\cite{abdulsalam:15}. Hence, being able to improve the performance of systems using AI would not only improve end user satisfaction, but also potentially reduce the energy footprint of future software.

To address this gap, we conducted an empirical study on LLM-driven optimization in large-scale open-source Java systems. We collected a dataset of 65 real performance-improving changes from four performance-sensitive Java projects. We developed an automated pipeline that tasks a set of LLMs with optimizing the original code. We test commercial models (OpenAI o4-mini\footnote{\label{fn:openai-04-mini}\url{https://openai.com/index/introducing-o3-and-o4-mini}} and Gemini 2.5 Pro\footnote{\label{fn:gemini-pro-2.5}\url{https://deepmind.google/models/gemini/pro/}}), as well as open-weight and reasoning models (DeepSeek-V3.2\footnote{\label{fn:deepseek-v3}\url{https://api-docs.deepseek.com/news/news251201}}, DeepSeek-R1-0528\footnote{\label{fn:deepseek-r1}\url{https://api-docs.deepseek.com/news/news250528}}).
We compared the LLM-generated code to both the original and the developer-improved versions of the code using developer-provided performance microbenchmarks written using the Java Microbenchmark Harness\footnote{\url{https://github.com/openjdk/jmh}\label{fn:jmh}} (JMH) to assess the performance of each version of the code. Our study answers the following research questions: \smallskip

\begin{summary}
\centering 
\textbf{RQ1: How effectively can LLMs optimize real-world software performance?}
\end{summary}

\noindent This research question investigates the capacity of LLMs to produce code that achieves substantial performance enhancements in real coding tasks (as opposed to, for example, online coding challenges). We systematically evaluate the impact of different prompt variations and model choices. Our results indicate that while LLMs exhibit a strong baseline capability---generating plausible patches in the majority of cases---they generally fall short of human-proposed performance improvements. Statistically, although LLM-generated solutions achieve large effect sizes compared to unoptimized code, human developers maintain a clear advantage across all tested configurations in three out of four assessed projects. In general, providing an explicit problem description in the prompt is the primary driver for unlocking optimization potential, whereas providing the code and no further instructions is often insufficient. Examining individual tasks reveals a more nuanced, and more volatile, reality. For some tasks, generated improvement suggestions actually cause performance regressions. However, in other tasks, LLMs are able to discover novel, non-obvious strategies that, in some cases, even outperform human developers.

\smallskip

\begin{summary}
\centering 
\textbf{RQ2: Do LLM-proposed optimization strategies align with human solutions?}
\end{summary}

\noindent Secondly, we investigate \emph{what kind} of performance fixes LLMs propose. We begin by analyzing the structural properties of the patches, finding that LLM-generated code differs significantly from developer solutions. It is more localized, yet consistently introduces higher Cyclomatic Complexity, implying a tendency to optimize via elaborate logic rather than structural simplification. While isolated instances of novel optimizations outperform human baselines (as noted in RQ1), performance gains remain highly dependent on mimicking human intuition. Providing explicit problem descriptions steers LLMs to solutions more similar to the developer baseline. However, even in these cases, actual implementations often tend to fall short. This underscores a critical limitation---while more problem context enables the model to accurately locate bottlenecks, it does not guarantee the synthesis of truly optimal code.


\smallskip Overall, our evaluation on real-world software reveals that LLM performance is highly volatile across different tasks. While models can surprisingly outperform humans in specific instances via novel strategies, they generally lag behind due to distinct structural complexity, marking them as promising yet inconsistent tools for practical software optimization.
Moreover, we conclude  that the current state of the art of evaluating
LLM code performance on algorithmic tasks~\cite{coignion:24, zheng2024beyond} provides an overly optimistic assessment of LLM capabilities. For real production performance, runtime information (e.g., profiling, tracing) matters. Future solutions need to be agent-based, incorporating not only static code features but also runtime information from profiling and benchmarking systems.



\section{Related Work}

\noindent\textbf{LLMs for Code Generation:} The application of LLMs to software engineering tasks has grown rapidly, building on the success of transformer architectures ~\cite{vaswani:2017}. Foundational models from OpenAI (GPT\footnote{\url{https://openai.com/index/introducing-gpt-5}}, o-series), Google (Gemini\footref{fn:gemini-pro-2.5}), Meta (Llama\footnote{\url{https://www.llama.com}}), and others (e.g., DeepSeek\footnote{\url{https://www.deepseek.com}} and Claude\footnote{\url{https://www.anthropic.com/claude}}) have demonstrated strong capabilities across the software development lifecycle, including requirement generation~\cite{nouri:24}, bug fixing~\cite{yang2024swe}, code refactoring~\cite{liu:25,shirafuji:23}, test case generation~\cite{chen:24,liu:25,codet:22}, and code understanding~\cite{nam:24,gu:24}. In parallel, research on automated program repair (APR)~\cite{apr:19} has increasingly adopted LLMs~\cite{xia:23,bouzenia:24}. Motivated by these advances, we construct an automated pipeline for patch generation with LLMs.

\smallskip\noindent\textbf{Benchmarking of AI-Generated Code:} The capabilities of these models are typically evaluated on two distinct tracks of benchmarks. The first track uses function- or file-level benchmarks, such as HumanEval~\cite{chen:21}, MBPP~\cite{austin:21}, APPS~\cite{hendrycks:21}, Code Contests~\cite{li2022competition}, and CoderEval~\cite{yu:24} to assess a model's ability to solve self-contained, algorithmic problems. Beyond correctness, recent research~\cite{zhong:24,aljedaani:24} has begun to explore additional quality dimensions of LLM-generated code, with a growing body of work focusing on performance efficiency. Some studies adapt function-level benchmarks to directly measure efficiency~\cite{liu2024evaluating,zheng2024beyond,peng:25} or to improve performance~\cite{shypula2023learning,PerfCodeGen:25}, while others use programming contest platforms like LeetCode to construct performance-oriented benchmarks~\cite{huang:24,coignion:24,qiu:25}. Recent work has also investigated incorporating execution traces to guide optimization, though with mixed results~\cite{di2025investigating}.

A key limitation of this track, however, is that success on small, isolated functions do not capture the complexity of working within large-scale, real-world codebases~\cite{han:16,jin:12}.
To address this, a second track has emerged for assessing functional correctness, using repository-level benchmarks, such as SWE-Bench~\cite{jimenez:23}, RepoBench~\cite{liu:23}, and RepoCoder~\cite{zhang:23}. These evaluate LLMs on their ability to resolve real GitHub issues within the full context of a software repository. These repository-level benchmarks have thus far focused exclusively on \emph{functional correctness}, that is, whether the generated patch resolves the issue and passes the existing test suite. Our work can be seen as a similar attempt to increase the level of realism in how we assess non-functional properties, specifically performance, of generated code, going from algorithmic tasks to more nuanced and varied real-world performance improvement tasks mined from real performance-sensitive projects.

\noindent\textbf{Performance Measurement:} Software performance engineering~\cite{smith1991performance} is a mature field with a long tradition of rigorous evaluation methodologies~\cite{sim:03}, supported by established tools such as Google Benchmark\footnote{\url{https://github.com/google/benchmark}}, Google Landmarks~\cite{weyand:20}, and SPEC\footnote{\url{https://www.spec.org/products}}. Among these, the Java Microbenchmark Harness (JMH) is a cornerstone for performance engineering in Java, and has delivered reliable and unbiased measurements as demonstrated in prior work~\cite{traini:23,laaber:20}. Earlier work on LLM performance assessment generally does not incorporate measurement best practices, working with unreliable tools such as timed test cases rather than real developer benchmarks. This is another way how our research improves on the level of realism and reliability in comparison to previous studies.


\section{Methodology}
\label{sec:method}

\begin{figure}[h!]
    \centering
    \includegraphics[width=1\linewidth]{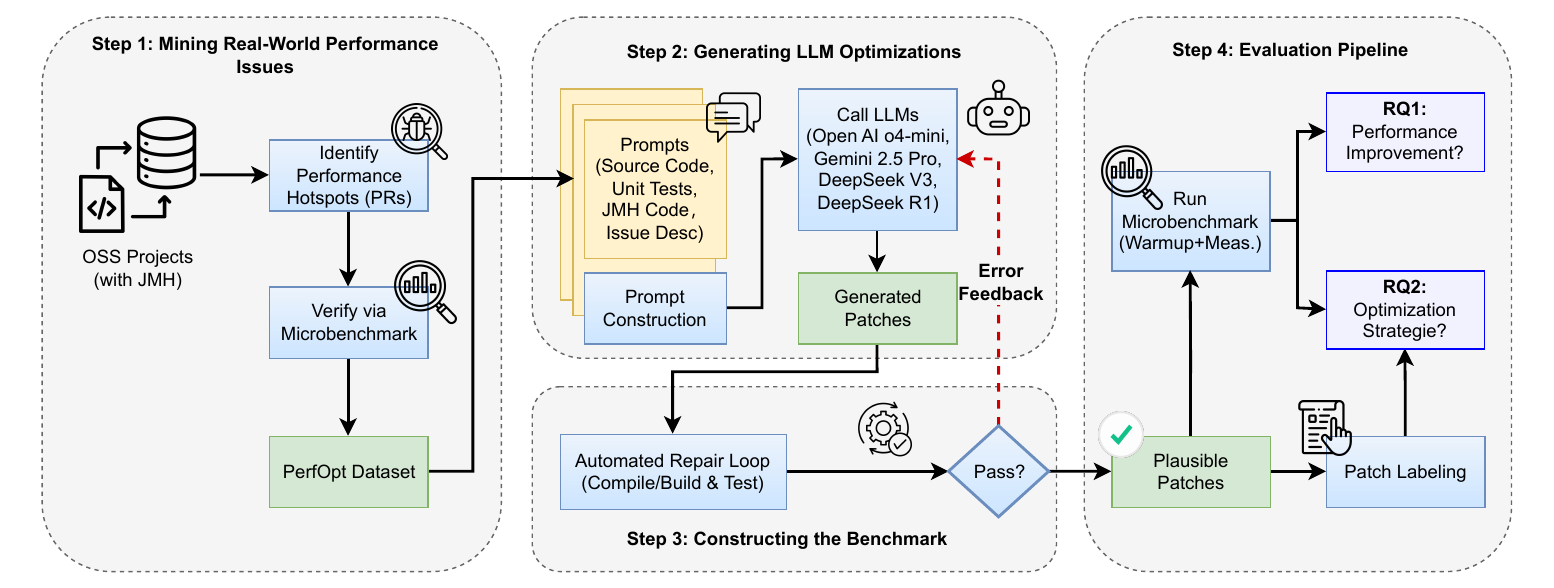}
    \caption{Overview of the research process.}
    \label{fig:workflow}
\end{figure}

We conduct our research in four steps, as shown in Figure~\ref{fig:workflow}. We begin by collecting a specialized dataset of real-world tasks where developers have achieved performance improvements, measuring the performance difference between the original and developer-improved code. Next, we configure four LLMs with four prompting strategies 
and instruct them to generate performance improvements---validating the syntactic and semantic correctness of these improvements through compilation and unit tests. We then benchmark the performance of the generated code using JMH microbenchmarks. Finally, we analyze and compare the nature and efficiency of LLM-generated patches against human-authored improvements to assess the practical potential of LLMs for performance improvement. In the following subsections, we discuss our methods in detail.

\subsection{Data Collection}
\label{sec:data-collection}

To construct a realistic and challenging evaluation suite, we specifically targeted Java due to its importance for high-performance enterprise systems and the availability of the Java Microbenchmark Harness (JMH), a mature framework for rigorous and fine-grained performance testing. Additionally, there is a plethora of Java-based open-source libraries with stringent performance requirements and substantial prior research on performance testing Java open source systems~\cite{laaber:20,traini:23,samoaa:21}. 
We curated the \texttt{PerfOpt Dataset}, a collection of 65 performance-oriented programming tasks mined from four open-source projects: Apache Kafka\footnote{\label{fn:kafka}\url{https://github.com/apache/kafka}}, Netty\footnote{\label{fn:netty}\url{https://github.com/netty/netty}}, Presto\footnote{\label{fn:presto}\url{https://github.com/prestodb/presto}}, and RoaringBitmap\footnote{\label{fn:rb}\url{https://github.com/RoaringBitmap/RoaringBitmap}}. The \texttt{PerfOpt} dataset is available in our online appendix\footref{fn:rp}.

\subsubsection{Data Collection Process}

It is important to note that constructing a reliable and realistic performance benchmark suite is significantly more challenging than traditional functional bug datasets---for our RQs, we required a high-quality dataset of issues that, rather than demonstrating behavior that violates the functional requirements, have demonstrable performance deficiencies. Concrete tasks were selected from projects via purposive sampling~\cite{baltes:22}. To ensure high quality, we screened more than 3000 commits among 30 open-source projects to initially identify 116 potential candidates that met our strict acceptance criteria: (1) the contribution was fixing a performance problem, (2) the contribution included at least one JMH benchmark to demonstrate the improvement, (3) running the benchmark on the original and improved version of the code showed a performance improvement, and (4), the code changes were concise, affecting no more than three files, to exclude large-scale refactoring (while still enabling assessment of difficult multi-file performance changes, as per our study goal).

Each commit meeting the criteria was then subjected to a two-step validation process: \textbf{functional verification} using unit tests and \textbf{performance validation} using JMH. After confirming a raw performance improvement, we formally determined whether that improvement was statistically significant. Each execution of a benchmark yields a set of performance measurements. A task is only included in the dataset if the performance difference for that set of observations between the original code and the developer-written solution is statistically significant according to a Wilcoxon signed-rank test~\cite{Wilcoxon45:mww} ($p < 0.05$) and demonstrates at least a medium effect size according to the Vargha-Delaney $\hat{A}_{12}$~\cite{Vargha00:Measure} measure ($\hat{A}_{12} \geq 0.64$). This resulted in the final set of 65 high-quality cases from four projects, where we know a performance improvement is possible, verified by developer benchmarks, and statistically replicable in our measurement infrastructure.

\subsubsection{PerfOpt Dataset}

We define a \emph{Task} in \texttt{PerfOpt} not merely as a function to be synthesized, but as a context-dependent optimization problem rooted in an existing codebase. Specifically, the granularity is \emph{multi-file}---the model is presented with the full content of the set of files relevant to the performance hotspot. Each task in \texttt{PerfOpt} captures a real-world optimization scenario involving issues such as algorithmic inefficiencies, inefficient memory allocations, or concurrency bottlenecks, distinguishing our work from isolated function- or single-file benchmarks (e.g., HumanEval). 

\begin{table}[h!]
\centering
\caption{Overview of the \texttt{PerfOpt} dataset: task counts and human performance baselines (LOC and Speedup).}
\label{tab:perfopt}
\scriptsize
\begin{tabular}{lc ccc ccc}
\toprule
\multirow{2}{*}{Project} & \multirow{2}{*}{\# Tasks} & \multicolumn{3}{c}{LOC Modified} & \multicolumn{3}{c}{Improvement} \\
\cmidrule(lr){3-5} \cmidrule(lr){6-8}
 & & Min & Max & Avg & Min & Max & Avg  \\
\midrule
\rowcolor{gray!15} Apache Kafka & 16 & 2 & 75 & 24.88 & 1.06 & 1401.47 & 99.14 \\
 Netty & 16 & 2 & 220 & 57.44 & 1.06 & 6.30 & 1.79 \\
 \rowcolor{gray!15} Presto & 27 & 2 & 127 & 31.74 & 1.05 & 145169.44 & 927.59 \\
 RoaringBitmap & 6 & 4 & 38 & 14.00 & 1.08 & 4.99 & 2.21 \\
\bottomrule
\end{tabular}
\end{table}

\noindent For each task, the dataset includes the following artifacts:
\begin{enumerate}
    \item \textbf{Performance Issue Description:} A natural-language description of the performance problem, extracted manually from developer commit messages or issue tracker reports.
    \item \textbf{Original Code:} The unoptimized source files.
    \item \textbf{Developer Solution:} The developer implementation that fixes the performance issue.
    \item \textbf{Unit Tests:} A test suite used to verify the plausibility of generated patches. This includes tests identified via naming conventions (e.g., \texttt{ClassNameTest}) and, where necessary, via dependency graph analysis to capture indirect invocations.
    \item \textbf{JMH Benchmarks:} The JMH benchmark code developers used to demonstrate their improvement (measuring throughput or execution time).
\end{enumerate}

We summarize the \texttt{PerfOpt} dataset in Table~\ref{tab:perfopt}. Reported performance improvements are for the developer patch, based on a benchmark score that will be detailed in Section~\ref{sec:analysis}.

\subsection{Generating LLM Optimizations}

\subsubsection{Model Selection} 
In this study, we evaluated a diverse set of LLMs to assess performance optimization capabilities across different architectures and licensing models. We selected OpenAI o4-mini (o4-mini) and Gemini 2.5 Pro (Gemini) as representatives of high-performance commercial models widely used by practitioners. Additionally, we included DeepSeek-V3 (DS-V3) and DeepSeek-R1 (DS-R1) to investigate the capability of state-of-the-art open-weight models, particularly focusing on the reasoning-enhanced capabilities of DS-R1. All models were accessed via their provider's API. We applied the settings $temperature = 0.3$ and $top\_p = 0.95$, selected via informal experimentation to balance creativity with adherence to syntax. For DeepSeek-R1, we also utilized its reasoning mode. As LLMs yield non-deterministic results, we repeated the generation process three times for each combination of task, prompting strategy, and model. We refer to these repetitions as \emph{trials}.

\subsubsection{Prompting Strategy}

\begin{figure}[htbp]
    \centering
    \includegraphics[width=0.7\linewidth]{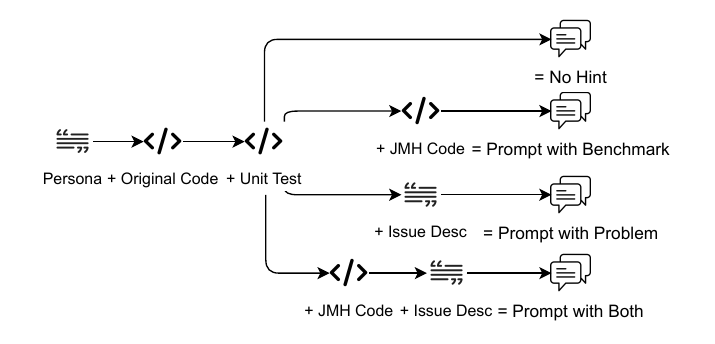}
    \caption{Overview of the prompt construction.}
    \label{fig:prompt_strategy}
\end{figure}

To assess the impact of varying forms of context on the performance task, we formulated four distinct prompt strategies, as illustrated in Figure~\ref{fig:prompt_strategy}.

\begin{itemize}
\item \textbf{No-Hint Prompt:} Contains the code and a general request to improve performance while preserving functionality. This measures the model's raw optimization capability without external guidance.

\item \textbf{Prompt with Benchmark:} Augments the baseline with the JMH benchmark source code. This provides structural pointers to the performance-critical paths and clarifies the evaluation metric.

\item \textbf{Prompt with Problem Description:} Augments the baseline with a curated natural-language summary of the performance issue. This tests the model's ability to leverage semantic clues without explicitly revealing the solution.

\item \textbf{Prompt with Both:} Combines both the benchmark code and problem description. This represents the comprehensive scenario where a bottleneck is both well-documented and benchmarked.
\end{itemize}

\noindent This tiered design enables a fine-grained analysis of how different types of context information affect the models' optimization effectiveness.

\subsubsection{Error-Guided Repair}
\label{sec:repair}

Our workflow integrates an automated, iterative repair loop to handle implausible responses, a concept central to the field of automated program repair (APR)~\cite{apr:19}. We mandate that models output changes as self-contained search-and-replace blocks in a structured JSON format, avoiding the fragility of line-number-based Git diffs which LLMs often struggle to generate accurately~\cite{ahn2024large,xu2025llm}.

Following the definition of \emph{plausibility} from APR literature~\cite{qi:15}, a response is considered \emph{implausible} if (a) the LLM did not produce a valid patch in the prescribed format, (b) the patched code failed to compile, or (c), the patched code did not pass the provided unit test suite. Conversely, a solution that meets all of these conditions is considered \textit{plausible}.

When we encountered an implausible patch, we automatically construct a new repair prompt containing the original prompt, the model's failed output, and the corresponding diagnostic message (e.g., the compilation error). This feedback loop enables the model to potentially self-correct implausible solutions. To avoid infinite looping, we end the repair loop after a maximum of \emph{two} attempts for integration failures and \emph{three} for verification failures. More attempts have not led to a successful repair in our experimentation. 

\subsection{JMH Execution}
We collected measurements for (a) the original code version, (b) the developer-patched version, and (c) each plausible LLM-generated patch in a controlled virtualized environment that our institution has access to\footnote{Information retracted for double-blind review.}. To ensure statistical rigor and reproducibility, we standardized the execution environment and protocols as follows:

\noindent \textbf{VM Resource Allocation:} We tailored computing resources to project requirements. The resource-intensive Presto project was assigned 4 vCPUs and 16 GB RAM, while other projects (Kafka, Netty, RoaringBitmap) were allocated 2 vCPUs and 8 GB RAM. All VMs were provisioned with 30--50 GB of disk space.

\noindent \textbf{JMH Configuration:} We adhered to the guidelines by Laaber et al.~\cite{laaber:20} to minimize measurement noise. Benchmarks were executed in a single fork with \textbf{10} warm-up and \textbf{50} measurement iterations (1s each). To mitigate the inherent variability of virtualized environments, we employed the \emph{Multiple Consecutive Trials} strategy~\cite{abedi:17}, ensuring statistical robustness against transient system noise.

In total, we collected a large dataset of 879,421 raw performance measurements as a basis for our analysis. Collecting this dataset required approximately 2799 VM hours, or 112 days, of raw computing time for collecting measurements.

\subsection{Analysis}
\label{sec:analysis}

Finally, we performed a rigorous analysis of the collected measurement data.

\subsubsection{Benchmark Scores}
JMH benchmarks report one measurement result for each iteration (50 in our case), and it is common to interpret the arithmetic mean result of all iterations as the (individual) benchmark score.\footnote{\url{https://stackoverflow.com/questions/24928922/jmh-what-does-the-score-value-mean}} However, JMH supports different ``modes'' (e.g., throughput, execution time), and not all benchmarks in \texttt{PerfOpt} use the same mode. Additionally, individual benchmark scores in our study vary by orders of magnitude. Finally, for our RQs, we are not inherently interested in the absolute benchmark scores, but in the \emph{change} in benchmark scores for the patched code in comparison to the original, unpatched version. 

Hence, we define a metric, \emph{patch performance score (pss)}, for a task \emph{id} as $pss_{id} = \frac{\mu_{id, new}}{\mu_{id, orig}}$ for benchmarks where higher values are better (such as throughput), and as $pss_{id} = \frac{\mu_{id, orig}}{\mu_{id, new}}$ for benchmarks where lower values are better (e.g., execution time). $\mu_{id, orig}$ and $\mu_{id, new}$ are the average performance score of the benchmark of the original and patched code, respectively. $pss$ is a ratio score in the interval $[0; \infty]$, with $pss = 1.0$ representing no performance change, $pss < 1.0$ representing a slowdown, and $pss > 1.0$ representing an improvement.

A further challenge in our analysis is that each task in \texttt{PerfOpt} may be associated with multiple benchmarks, and benchmarks may have multiple configurations (for practical purposes it is common to consider each distinct benchmark configuration an entirely separate benchmark~\cite{samoaa:21}). Not all benchmarks will profit from an optimization, and developers often also refer to benchmarks that are not expected to change in their commits (e.g., as baselines or guardrails). Hence, we select a single benchmark and configuration for each task to serve as the basis for our analysis, referred to as the \emph{representative benchmark}. The representative benchmark is defined as the specific benchmark configuration where the human developer's patch achieved the highest $pss$. The performance of all other solutions (including all LLM trials) for that task is then measured and reported using their results in the representative benchmark. This is based on the intuition that this benchmark best exemplifies the improvements that the developers intended to achieve by changing the code.



\subsubsection{Statistical Analysis}\label{sec:stat}

To answer RQ1, we first selected a single \emph{representative LLM solution} from the three trials for each model-prompt combination. The representative solution is defined as follows: (a) if all three trials lead to plausible patches, we use the solution with median $pss$, (b) if two patches are plausible, we conservatively use the worse-performing patch, and (c) if only one patch is plausible, we use that patch.

We then conducted a pairwise statistical comparison using the $pss$ from the representative LLM solution and the $pss$ of the developer solution (or the unmodified baseline). We employed a two-step statistical procedure. First, we used a two-sided Wilcoxon signed-rank test~\cite{Wilcoxon45:mww} to determine whether the performance scores of two patches are drawn from different distributions. The null hypothesis ($H_0$) is that the scores are drawn from the same distribution. If the resulting p-value is less than 0.05, we reject $H_0$. Second, for all pairs with a statistically significant difference, we calculated the Vargha-Delaney $\hat{A}_{12}$ effect size~\cite{Vargha00:Measure}. This non-parametric statistic provides a probabilistic measure of the difference, where a value of $\hat{A}_{12}(X, Y)$ represents the probability that a random sample from group $X$ is greater than a random sample from group $Y$. An $\hat{A}_{12}$ value of 0.5 indicates stochastic equality, while a value of $0.50 - 0.55$ indicates a negligible effect, $0.56 - 0.63$ indicates a small effect, $0.64 - 0.70$ indicates a medium effect, and $\geq 0.71$ indicates a large effect. Values below $0.50$ indicate that $X < Y$, with effect sizes at the same ranges in the opposite direction.

\subsubsection{Manual Patch labeling}
\label{sec:labeling}
To answer RQ2, we labeled all LLM-generated patches with regard to how similar they are to the improvement proposed by human developers. We use the following pre-defined codes:

\begin{itemize}
\item \textbf{Strategy Match:} The patch targets the same bottleneck and employs an identical or nearly identical optimization logic to the developer's solution.
\item \textbf{Strategy Alignment:} The patch correctly targets the same performance bottleneck as the developer but employs a distinct algorithmic approach or implementation to resolve it.
\item \textbf{Strategy Divergence:} The patch targets an unrelated code region or attempts to optimize a different performance issue, missing the developer's intended optimization target.
\end{itemize}

Each generated patch is assigned exactly one of these codes by the first author. To ensure reliability, we conducted an inter-rater reliability assessment on 40 randomly-sampled patches, labeled by an independent evaluator to validate the first author's classification of the full dataset. We calculated Cohen's Kappa ($\kappa$) to measure agreement. The resulting $\kappa$ value was 0.77, indicating \emph{substantial agreement}~\cite{landis1977measurement}, with an exact agreement rate of 85\%. The analysis revealed that \emph{Strategy Divergence} was highly distinguishable (nearly 100\% agreement). The primary source of disagreement lay in the nuance between \emph{Strategy Alignment} and \emph{Strategy Match}, where the independent evaluator occasionally applied a slightly stricter threshold for algorithmic identity. Given the substantial agreement level, the first author's classifications were retained for the analysis.

\section{Results}
We now present the results of our study. We begin by analyzing the plausibility of patches generated by the LLMs (Section~\ref{sec:plausibility}), which serves as the foundation for evaluating their performance effectiveness (RQ1, Section~\ref{sec:rq1_results}) and similarity to human optimization strategies (RQ2, Section~\ref{sec:rq3_results}).

\subsection{Patch Generation and Plausibility}\label{sec:plausibility}

Before assessing the extent to which LLMs improve performance, we first must understand whether LLM-generated solutions are plausible as per the definition introduced in Section~\ref{sec:repair}. Table~\ref{tab:patch_count} provides a detailed breakdown of the generated solutions for each model and prompting strategy. To decouple the model's inherent coding capability from its ability to repair implausible patches, we divide assessment into (1) the initial success rate ($SR_{init}$), which measures the proportion of tasks where the model's first attempt passed all tests without requiring repair, and (2), the repair success rate ($SR_{repair}$), which calculates the percentage of \emph{initially failed} solutions that were successfully fixed within the allowed repair iterations.

\begin{table}[h!]
\centering
\caption{Overview of patch plausibility by model and prompt strategy. $SR_{init}$ denotes the success rate of the first attempt, while $SR_{repair}$ measures the repair rate calculated over the initially implausible subset.}
\label{tab:patch_count}
\scriptsize
\begin{tabular}{l cccc c | cc}
\toprule
\multirow{2}{*}{\textbf{Model}} & \multicolumn{4}{c}{\textbf{Prompt Strategy}} & \multirow{2}{*}{\textbf{Total}} & \multicolumn{2}{c}{\textbf{Repair Breakdown}} \\
\cmidrule(lr){2-5} \cmidrule(lr){7-8}
 & {No Hint} & {Bench.} & {Problem} & {Both} & &{$SR_{init}$} & {$SR_{repair}$} \\
\midrule
OpenAI o4-mini & 116 (59\%) & 109 (56\%) & 103 (53\%) & 117 (60\%) & \textbf{445 (57\%)}  & 356 (47\%) & {89 (21\%)} \\
Gemini 2.5 Pro & 126 (65\%) & 134 (69\%) & 129 (66\%) & 130 (67\%) & \textbf{519 (67\%)}  & 389 (50\%) & {130 (33\%)} \\
DeepSeek V3 & 118 (61\%) & 124 (64\%) & 120 (62\%) & 125 (64\%) & \textbf{487 (62\%)} & 386 (49\%) & 101 (27\%) \\
DeepSeek R1 & 120 (62\%) & 131 (67\%) & 122 (43\%) & 129 (66\%) & \textbf{502 (64\%)} & 424 (54\%) & 78 (22\%) \\
\bottomrule
\end{tabular}
\end{table}

As shown in the \textit{Total} column of Table~\ref{tab:patch_count}, the models achieve overall plausibility rates ranging from 57\% (OpenAI o4-mini) to 67\% (Gemini 2.5 Pro). These figures notably trail the high Pass@1 rates typically reported in algorithmic benchmarks like COFFE~\cite{peng:25}. Crucially, this performance gap persists even though we employ significantly more capable models (e.g., Gemini 2.5 Pro vs. Gemini 1.5 Pro), underscoring the inherent complexity of real-world performance engineering compared to isolated algorithmic tasks, as used in prior research.

A critical insight from Table~\ref{tab:patch_count} lies in the repair breakdown. R1 is more likely to generate correct code on the first attempt compared to other models, achieving an $SR_{init}$ of 54\%. Conversely, Gemini 2.5 Pro demonstrates superior repair capabilities. Although its initial success rate (50\%) is lower than R1's, it achieves a higher $SR_{repair}\%$ (vs. 22\% for R1) and, consequently, the best overall success rate. This comparison reveals distinct model behaviors---while DeepSeek R1 relies on its strong reasoning capability to minimize initial errors, Gemini effectively leverages the automated repair loop to salvage initially invalid solutions. OpenAI o4-mini, with the lowest repair rate (21\%), appears less effective at utilizing feedback to fix compilation or logic errors.

\subsection{Can LLMs Optimize Real-World System Performance? (RQ1)}\label{sec:rq1_results}

To answer RQ1, we assess whether the generated plausible patches actually improve performance. 

\subsubsection{Aggregate Performance Analysis}

\begin{figure}[h!]
    \centering
    \includegraphics[width=\linewidth]{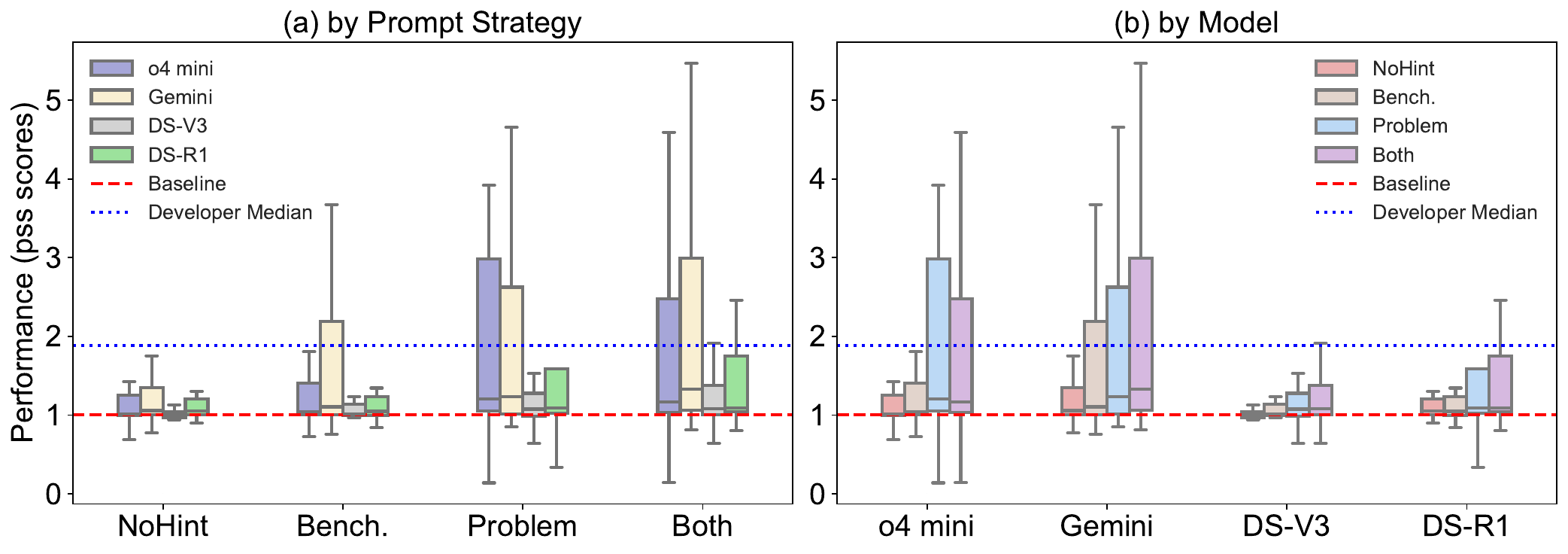}
    \caption{Distribution of $pss$ scores for LLM-generated solutions.}
    \label{fig:rq1}
\end{figure}

Figure~\ref{fig:rq1} depicts the distribution of $pss$ scores grouped by prompting strategy (left) and by model (right). For visual clarity in the boxplots, outliers are not displayed, where an outlier is defined using the interquartile range~\cite{wan2014estimating}. Most generated solutions outperform the baseline ($pss>1$), confirming that LLMs can generally improve code performance. Gemini offers the largest median improvements (but also the highest variance, particularly when provided with full context). In contrast, DeepSeek-V3 exhibits a remarkably conservative distribution---its improvements cluster tightly around the baseline across all prompting strategies, suggesting a tendency to preserve the original logic over aggressive optimization. DeepSeek-R1 occupies a middle ground, showing consistent but moderate improvements, with less volatility than o4-mini or Gemini.

Analyzing prompting strategies reveals further divergences. If no additional context is provided in the prompt (no hint), all models tend to produce only incremental improvements. For o4-mini, providing a concise problem description is sometimes extremely valuable, indicated by a higher third quartile. Providing a combination of both types of information sometimes leads to a decrease in performance (as also reported by Khojah et al.~\cite{khojah:25}). However, Gemini and DeepSeek show clear improvements under the ``both'' strategy, achieving their highest median performance.



\begin{figure}[h!]
    \centering
    \begin{minipage}[b]{0.60\textwidth}
        \centering
        \begin{tikzpicture}
            \begin{axis}[
                width=\linewidth,
                height=6.6cm,
                xbar, 
                xmin=0, xmax=1,
                ymin=0.5, ymax=16.5,
                axis on top,
                xtick={0, 0.29, 0.36, 0.44, 0.5, 0.56, 0.64, 0.71, 1},
                xticklabels={0, 0.29, 0.36, 0.44, 0.50, 0.56, 0.64, 0.71, 1},
                xticklabel style={font=\scriptsize, rotate=45, anchor=east},
                ytick={1,2,3,4,5,6,7,8,9,10,11,12,13,14,15,16},
                yticklabels={
                    {o4-mini (NoHint)}, {o4-mini (Bench)}, {o4-mini (Prob)}, {o4-mini (Both)},
                    {Gemini (NoHint)}, {Gemini (Bench)}, {Gemini (Prob)}, {Gemini (Both)},
                    {DS-V3 (NoHint)}, {DS-V3 (Bench)}, {DS-V3 (Prob)}, {DS-V3 (Both)},
                    {DS-R1 (NoHint)}, {DS-R1 (Bench)}, {DS-R1 (Prob)}, {DS-R1 (Both)}
                },
                yticklabel style={font=\scriptsize},
                y dir=reverse, 
                legend style={
                    at={(0.5,1.08)}, 
                    anchor=south, 
                    legend columns=4, 
                    font=\scriptsize,
                    draw=none, 
                    column sep=0.4cm, 
                },
                grid=major,
                title={(a) Effect Size by Model Configuration},
                title style={at={(0.5,-0.15)}, anchor=north, font=\small}
            ]
            \fill[black!12] (axis cs:0, \pgfkeysvalueof{/pgfplots/ymin}) rectangle (axis cs:0.29, \pgfkeysvalueof{/pgfplots/ymax});
            \fill[black!12] (axis cs:0.71, \pgfkeysvalueof{/pgfplots/ymin}) rectangle (axis cs:1, \pgfkeysvalueof{/pgfplots/ymax});
            \fill[black!08] (axis cs:0.29, \pgfkeysvalueof{/pgfplots/ymin}) rectangle (axis cs:0.36, \pgfkeysvalueof{/pgfplots/ymax});
            \fill[black!08] (axis cs:0.64, \pgfkeysvalueof{/pgfplots/ymin}) rectangle (axis cs:0.71, \pgfkeysvalueof{/pgfplots/ymax});
            \fill[black!04] (axis cs:0.36, \pgfkeysvalueof{/pgfplots/ymin}) rectangle (axis cs:0.44, \pgfkeysvalueof{/pgfplots/ymax});
            \fill[black!04] (axis cs:0.56, \pgfkeysvalueof{/pgfplots/ymin}) rectangle (axis cs:0.64, \pgfkeysvalueof{/pgfplots/ymax});

            \draw[black, thick, dashed] (axis cs:0.5, 0) -- (axis cs:0.5, 17);
            \node[fill=white, inner sep=1pt, text=gray] at (axis cs:0.5, 17) {};

            \def\colorMini{violet!70!white}
            \def\colorGemini{green!60!black}
            \def\colorVthree{orange!80!black}
            \def\colorRone{blue!70!black}

            \addlegendimage{no markers, fill=\colorMini, area legend}\addlegendentry{\textcolor{\colorMini}{o4-mini}}
            \addlegendimage{no markers, fill=\colorGemini, area legend}\addlegendentry{\textcolor{\colorGemini}{Gemini}}
            \addlegendimage{no markers, fill=\colorVthree, area legend}\addlegendentry{\textcolor{\colorVthree}{DS-V3}}
            \addlegendimage{no markers, fill=\colorRone, area legend}\addlegendentry{\textcolor{\colorRone}{DS-R1}}
            \addlegendimage{empty legend}\addlegendentry{}
            \addlegendimage{only marks, mark=triangle*, color=gray}\addlegendentry{\textbf{vs. Developer}}
            \addlegendimage{only marks, mark=*, color=gray}\addlegendentry{\textbf{vs. Original}}
            \addlegendimage{empty legend}\addlegendentry{}
            \addplot[only marks, mark=*, mark size=2.5pt, color=\colorMini] coordinates { (0.875, 3) (0.689, 2) (0.897, 4) };
            \addplot[only marks, mark=triangle*, mark size=3pt, color=\colorMini] coordinates { (0.142, 1) (0.320, 3) (0.192, 2) (0.317, 4) };
            
            \addplot[only marks, mark=*, mark size=2.5pt, color=\colorGemini] coordinates { (0.679, 5) (0.885, 7) (0.812, 6) (0.888, 8) };
            \addplot[only marks, mark=triangle*, mark size=3pt, color=\colorGemini] coordinates { (0.201, 5) (0.305, 7) (0.270, 6) (0.351, 8) };
            
            \addplot[only marks, mark=*, mark size=2.5pt, color=\colorVthree] coordinates { (0.646, 11) (0.661, 10) (0.750, 12) };
            \addplot[only marks, mark=triangle*, mark size=3pt, color=\colorVthree] coordinates { (0.076, 9) (0.207, 11) (0.168, 10) (0.232, 12) };
            
            \addplot[only marks, mark=*, mark size=2.5pt, color=\colorRone] coordinates { (0.750, 13) (0.806, 15) (0.712, 14) (0.786, 16) };
            \addplot[only marks, mark=triangle*, mark size=3pt, color=\colorRone] coordinates { (0.170, 13) (0.265, 15) (0.206, 14) (0.302, 16) };
            \end{axis}
        \end{tikzpicture}
    \end{minipage}%
    \hfill
    \begin{minipage}[b]{0.33\textwidth} 
        \centering
        \begin{tikzpicture}
            \begin{axis}[
                scale only axis,
                width=0.76\linewidth, 
                height=0.76\linewidth,
                axis on top,
                xmin=-0.5, xmax=3.5,
                ymin=-0.5, ymax=3.5,
                xtick={0, 1, 2, 3},
                xticklabels={NoHint, Bench, Prob, Both},
                xticklabel style={font=\scriptsize, rotate=45, anchor=north east},
                ytick={0, 1, 2, 3},
                yticklabels={Both, Prob, Bench, NoHint}, 
                yticklabel style={font=\scriptsize},
                colorbar horizontal,
                colorbar style={
                    at={(0.5, 1.06)},
                    anchor=south, 
                    width=0.76\linewidth,
                    height=0.15cm,
                    xtick={0, 0.5, 1},
                    xticklabel pos=upper,
                    xticklabel style={
                        font=\tiny,
                        yshift=1pt,
                    }
                },
                colormap name=bluewhitered,
                title={(b) Prompt Effect Size},
                title style={at={(0.5,-0.3)}, anchor=north, font=\small},
                nodes near coords={
                    \pgfmathfloatparsenumber{\pgfplotspointmeta}%
                    \pgfmathfloatgetflagstomacro\pgfmathresult\flags%
                    \ifnum\flags=3\relax%
                    \else%
                        \pgfmathprintnumber[fixed, precision=2]{\pgfplotspointmeta}%
                    \fi%
                },
                nodes near coords style={
                    font=\tiny, 
                    color=black, 
                    yshift=-1pt 
                },
                nodes near coords align=center,
                enlargelimits=false,
                unbounded coords=jump
            ]
            
            \addplot[
                matrix plot*,
                mesh/cols=4,
                point meta=explicit,
            ] coordinates {
                (0,0) [0.68]  
                (1,0) [0.61]  
                (2,0) [nan]   
                (3,0) [nan]   
                
                (0,1) [0.66]  
                (1,1) [nan]   
                (2,1) [nan]   
                (3,1) [nan]   
                
                (0,2) [0.56]  
                (1,2) [nan]   
                (2,2) [nan]   
                (3,2) [0.39]  
                
                (0,3) [nan]   
                (1,3) [0.44]  
                (2,3) [0.34]  
                (3,3) [0.32]  
            };
            \end{axis}
        \end{tikzpicture}
    \end{minipage}

    \caption{Vargha-Delaney $\hat{A}_{12}$ effect size. Effect sizes are shown only for pairs with a statistically significant difference ($p < 0.05$ from Wilcoxon signed-rank test). \textbf{(a)} Comparison of individual model configurations against Original (circles) and Developer (triangles) baselines. Background bands indicate effect size magnitude: Negligible (white), Small (light gray), Medium (gray), and Large (dark gray). \textbf{(b)} Pairwise effect size heatmap of prompting strategies. White fields indicate no statistically significant difference.}
    \label{fig:effect_size}
\end{figure}

Figure~\ref{fig:effect_size} presents the results of a statistical evaluation. The magnitude of the Vargha-Delaney $\hat{A}_{12}$ effect sizes is interpreted based on the thresholds defined in Section~\ref{sec:stat}. Figure~\ref{fig:effect_size}(a) details the performance of each LLM configuration compared to the baselines. We observe that all LLM configurations achieve performance improvements over the original code, except for o4-mini and DS-V3 with the ``NoHint'' prompt, which did not achieve a statistically significant improvement over the original baseline. As shown by the circle markers, most configurations yield $\hat{A}_{12}$ values exceeding the $0.71$ threshold, indicating a large effect size.  However, developer solutions (triangles) remain superior to every LLM configuration. The $\hat{A}_{12}$ values for LLMs against developers cluster primarily between $0.15$ and $0.35$, representing a medium to large effect size in favor of human experts. This shows that LLMs are, indeed, effective at generating performance-enhancing patches, but do not reach human-competitiveness (although exceptions exist).

Figure~\ref{fig:effect_size}(b) shows the pairwise statistical comparison of prompting strategies, aggregated across models. In this heatmap, colored cells indicate a statistically significant difference ($p < 0.05$), while empty cells indicate no significant difference. Analyzing this heatmap reveals a clear hierarchy. Patches generated without context (``NoHint'') are significantly outperformed by all others. Crucially, the empty cell between ``Problem'' and ``Both'' indicates no statistically significant difference, suggesting that providing a clear problem description is the primary driver for performance optimization. While providing benchmark code improves over ``NoHint'', it is significantly less effective than the combined approach (``Both''), indicating that code context alone is insufficient without semantic guidance.

\subsubsection{Task-Specific Performance Analysis}\label{Task-Specific}

\begin{table}[h!]
\centering
\scriptsize
\caption{Task-specific performance comparison of $pss$ scores between a representative commercial model ({Gemini 2.5 Pro}) and a reasoning-enhanced open-weight model ({DeepSeek R1}). Results are shown for the representative trial, grouped by project and sorted by developer improvement magnitude. A \colorbox{lightgreen}{green background} indicates that the LLM solution outperformed the developer. A \colorbox{lightred}{red background} indicates a performance regression. Note that 21 tasks where neither model was able to generate a plausible solution are omitted.}
\label{tab:hash_performance_summary}
\resizebox{0.9\textwidth}{!}{%
\begin{tabular}{@{}c l l cccc cccc@{}}
\toprule
& \multicolumn{2}{c}{\textbf{ }} & \multicolumn{4}{c}{\textbf{Gemini 2.5 Pro}} & \multicolumn{4}{c}{\textbf{DeepSeek R1}} \\
\cmidrule(lr){4-7} \cmidrule(lr){8-11}
\textbf{Project} & \textbf{Hash} & \textbf{Dev} & NH & Bench & Prob & Both & NH & Bench & Prob & Both \\
\midrule
\midrule
\multirow{11}{*}{\rotatebox[origin=c]{90}{\textbf{Kafka}}} & \href{https://github.com/apache/kafka/commit/b25c96a}{b25c96a} & 1401.47 & \cellcolor{lightgreen}{1898.94} & \cellcolor{lightgreen}{1905.23} & \cellcolor{lightgreen}{1879.81} & \cellcolor{lightgreen}{1901.39} & - & \cellcolor{lightgreen}{1932.48} & \cellcolor{lightgreen}{1946.33} & \cellcolor{lightgreen}{1918.99} \\
 & \href{https://github.com/apache/kafka/commit/9bb2f78}{9bb2f78} & 654.58 & 647.27 & 549.45 & 654.25 & 632.65 & 534.61 & \cellcolor{lightgreen}{663.91} & \cellcolor{lightgreen}{664.91} & 634.21 \\
 & \href{https://github.com/apache/kafka/commit/4fa5bdc}{4fa5bdc} & 4.54 & 3.95 & 4.34 & 4.10 & 3.79 & 1.10 & - & 1.03 & 1.05 \\
 & \href{https://github.com/apache/kafka/commit/cfc34ca}{cfc34ca} & 3.95 & - & 3.68 & - & 1.83 & - & - & - & - \\
 & \href{https://github.com/apache/kafka/commit/e3ccf20}{e3ccf20} & 2.81 & \cellcolor{lightred}{0.96} & \cellcolor{lightred}{0.93} & \cellcolor{lightgreen}{3.45} & \cellcolor{lightgreen}{4.12} & - & - & \cellcolor{lightred}{0.39} & \cellcolor{lightred}{0.91} \\
 & \href{https://github.com/apache/kafka/commit/c8af740}{c8af740} & 2.10 & \cellcolor{lightgreen}{2.14} & \cellcolor{lightgreen}{2.19} & \cellcolor{lightgreen}{2.18} & \cellcolor{lightgreen}{2.19} & - & 1.13 & 2.07 & - \\
 & \href{https://github.com/apache/kafka/commit/e2e2c62}{e2e2c62} & 2.04 & 1.66 & - & - & 1.06 & - & \cellcolor{lightred}{0.97} & - & - \\
 & \href{https://github.com/apache/kafka/commit/c05403f}{c05403f} & 1.55 & 1.24 & 1.24 & 1.52 & 1.24 & 1.30 & 1.23 & 1.29 & 1.54 \\
 & \href{https://github.com/apache/kafka/commit/922a95a}{922a95a} & 1.28 & 1.14 & \cellcolor{lightgreen}{8.81} & 1.26 & 1.26 & - & - & - & 1.20 \\
 & \href{https://github.com/apache/kafka/commit/aecd47b}{aecd47b} & 1.05 & \cellcolor{lightgreen}{2.42} & \cellcolor{lightgreen}{2.30} & \cellcolor{lightgreen}{1.06} & \cellcolor{lightgreen}{1.06} & \cellcolor{lightgreen}{2.58} & \cellcolor{lightgreen}{2.43} & \cellcolor{lightgreen}{1.05} & 1.05 \\
\midrule
 & Avg & 7.27 & 5.96 & \cellcolor{lightgreen}{7.81} & 6.14 & 6.49 & 2.14 & 4.60 & 4.13 & 4.32 \\
\midrule
\multirow{11}{*}{\rotatebox[origin=c]{90}{\textbf{Netty}}} & \href{https://github.com/netty/netty/commit/2ed95c9}{2ed95c9} & 6.30 & 1.05 & 1.89 & 1.12 & \cellcolor{lightred}{0.81} & \cellcolor{lightred}{0.99} & - & 2.92 & - \\
 & \href{https://github.com/netty/netty/commit/87ec2f8}{87ec2f8} & 3.00 & {1.00} & 1.01 & 2.50 & 2.52 & - & - & 1.11 & 2.46 \\
 & \href{https://github.com/netty/netty/commit/b2eaab0}{b2eaab0} & 1.80 & 1.00 & 1.00 & 1.00 & 1.52 & 1.03 & 1.10 & 1.05 & 1.04 \\
 & \href{https://github.com/netty/netty/commit/83a19d5}{83a19d5} & 1.71 & 1.35 & \cellcolor{lightred}{0.83} & 1.28 & 1.24 & \cellcolor{lightred}{0.68} & \cellcolor{lightred}{0.61} & \cellcolor{lightred}{0.59} & \cellcolor{lightred}{0.75} \\
 & \href{https://github.com/netty/netty/commit/2791f0f}{2791f0f} & 1.45 & 1.01 & 1.03 & 1.37 & 1.42 & - & - & 1.18 & \cellcolor{lightgreen}{1.47} \\
 & \href{https://github.com/netty/netty/commit/67d3a78}{67d3a78} & 1.25 & \cellcolor{lightred}{0.99} & 1.07 & 1.22 & 1.02 & - & \cellcolor{lightred}{0.91} & \cellcolor{lightgreen}{1.28} & \cellcolor{lightgreen}{1.25} \\
 & \href{https://github.com/netty/netty/commit/deea51e}{deea51e} & 1.25 & {1.00} & 1.13 & 1.11 & 1.25 & - & - & - & - \\
 & \href{https://github.com/netty/netty/commit/5a9b28a}{5a9b28a} & 1.21 & \cellcolor{lightred}{0.95} & 1.14 & 1.09 & 1.08 & \cellcolor{lightred}{0.38} & \cellcolor{lightred}{0.58} & \cellcolor{lightred}{0.34} & \cellcolor{lightred}{0.80} \\
 & \href{https://github.com/netty/netty/commit/8a8337e}{8a8337e} & 1.19 & 1.06 & 1.04 & 1.11 & 1.01 & 1.04 & 1.14 & \cellcolor{lightgreen}{1.30} & \cellcolor{lightgreen}{1.21} \\
 & \href{https://github.com/netty/netty/commit/feb804d}{feb804d} & 1.10 & 1.07 & 1.09 & 1.08 & \cellcolor{lightgreen}{1.11} & \cellcolor{lightred}{0.67} & - & \cellcolor{lightred}{0.65} & \cellcolor{lightred}{0.69} \\
\midrule
 & Avg & 1.71 & 1.04 & 1.10 & 1.24 & 1.24 & \cellcolor{lightred}{0.84} & \cellcolor{lightred}{0.91} & \cellcolor{lightred}{0.98} & 1.09 \\
\midrule
\multirow{20}{*}{\rotatebox[origin=c]{90}{\textbf{Presto}}} & \href{https://github.com/prestodb/presto/commit/bfb18d6}{bfb18d6} & 145169.44 & 1.03 & 1.03 & 1.03 & 117653.53 & 1.03 & 1.03 & 1.04 & - \\
 & \href{https://github.com/prestodb/presto/commit/20dcaed}{20dcaed} & 2142.71 & 2119.50 & 2126.92 & 2129.77 & 2136.70 & - & - & - & - \\
 & \href{https://github.com/prestodb/presto/commit/239f726}{239f726} & 1033.72 & {1.00} & \cellcolor{lightgreen}{1087.83} & {1.00} & 1.03 & - & 1.02 & \cellcolor{lightgreen}{1060.24} & \cellcolor{lightgreen}{1087.03} \\
 & \href{https://github.com/prestodb/presto/commit/8d7f587}{8d7f587} & 136.32 & - & 1.01 & 130.84 & 132.86 & - & - & - & - \\
 & \href{https://github.com/prestodb/presto/commit/ee74e4b}{ee74e4b} & 38.54 & 1.00 & 1.00 & 1.00 & - & - & \cellcolor{lightred}{0.99} & \cellcolor{lightred}{0.99} & \cellcolor{lightred}{0.99} \\
 & \href{https://github.com/prestodb/presto/commit/8abba9e}{8abba9e} & 30.27 & - & 30.05 & - & - & - & - & - & - \\
 & \href{https://github.com/prestodb/presto/commit/3f63ade}{3f63ade} & 12.86 & \cellcolor{lightred}{0.98} & \cellcolor{lightred}{0.93} & 7.27 & 8.41 & 1.03 & - & - & - \\
 & \href{https://github.com/prestodb/presto/commit/46d219f}{46d219f} & 10.12 & \cellcolor{lightred}{0.97} & 1.01 & 10.03 & 10.02 & - & - & - & - \\
 & \href{https://github.com/prestodb/presto/commit/a26f440}{a26f440} & 9.66 & \cellcolor{lightred}{0.98} & \cellcolor{lightred}{0.89} & {1.00} & \cellcolor{lightred}{0.99} & \cellcolor{lightred}{0.96} & \cellcolor{lightred}{0.96} & - & \cellcolor{lightred}{0.95} \\
 & \href{https://github.com/prestodb/presto/commit/4e71bf3}{4e71bf3} & 2.45 & 1.75 & 1.83 & 2.27 & 2.33 & 1.19 & 1.02 & - & 1.94 \\
 & \href{https://github.com/prestodb/presto/commit/3a4f2e3}{3a4f2e3} & 1.89 & {1.00} & 1.00 & 1.10 & 1.11 & 1.21 & 1.02 & 1.05 & 1.07 \\
 & \href{https://github.com/prestodb/presto/commit/6908344}{6908344} & 1.87 & \cellcolor{lightred}{0.97} & - & 1.86 & \cellcolor{lightgreen}{2.22} & - & - & - & - \\
 & \href{https://github.com/prestodb/presto/commit/9e010a7}{9e010a7} & 1.80 & \cellcolor{lightred}{0.98} & \cellcolor{lightred}{0.92} & \cellcolor{lightred}{0.87} & \cellcolor{lightred}{0.87} & \cellcolor{lightred}{0.90} & \cellcolor{lightred}{0.74} & \cellcolor{lightred}{0.74} & \cellcolor{lightred}{0.77} \\
 & \href{https://github.com/prestodb/presto/commit/53aa7f6}{53aa7f6} & 1.40 & 1.01 & 1.00 & 1.00 & 1.00 & 1.00 & 1.00 & 1.01 & \cellcolor{lightred}{0.99} \\
 & \href{https://github.com/prestodb/presto/commit/b54d248}{b54d248} & 1.29 & 1.20 & 1.26 & 1.26 & 1.23 & 1.21 & 1.23 & 1.09 & - \\
 & \href{https://github.com/prestodb/presto/commit/efd0834}{efd0834} & 1.18 & - & - & {1.00} & {1.00} & - & - & - & \cellcolor{lightred}{0.98} \\
 & \href{https://github.com/prestodb/presto/commit/c3cb7a9}{c3cb7a9} & 1.10 & 1.00 & 1.10 & 1.01 & {1.00} & \cellcolor{lightred}{0.96} & \cellcolor{lightred}{0.99} & 1.09 & \cellcolor{lightred}{0.98} \\
 & \href{https://github.com/prestodb/presto/commit/297b089}{297b089} & 1.07 & 1.04 & 1.00 & 1.02 & 1.04 & 1.06 & 1.00 & 1.05 & {1.07} \\
 & \href{https://github.com/prestodb/presto/commit/659c3ee}{659c3ee} & 1.06 & \cellcolor{lightgreen}{1.08} & 1.01 & - & - & 1.05 & 1.05 & \cellcolor{lightred}{0.99} & \cellcolor{lightgreen}{1.07} \\
 \midrule
 & Avg & 13.42 & 1.56 & 2.69 & 2.65 & 4.99 & 1.03 & 1.00 & 1.44 & 1.48 \\
\midrule
\multirow{6}{*}{\rotatebox[origin=c]{90}{\textbf{RB}}} & \href{https://github.com/RoaringBitmap/RoaringBitmap/commit/34bce1b}{34bce1b} & 4.99 & 4.07 & 4.01 & 3.88 & 3.62 & 1.72 & 4.28 & \cellcolor{lightgreen}{5.16} & 2.39 \\
 & \href{https://github.com/RoaringBitmap/RoaringBitmap/commit/53ba593}{53ba593} & 2.14 & 1.30 & 1.34 & 1.00 & 1.97 & 1.00 & 1.34 & \cellcolor{lightred}{0.93} & \cellcolor{lightred}{0.93} \\
 & \href{https://github.com/RoaringBitmap/RoaringBitmap/commit/b5b4af6}{b5b4af6} & 1.94 & 1.93 & 1.32 & 1.14 & 1.92 & \cellcolor{lightred}{0.99} & - & 1.00 & \cellcolor{lightred}{0.89} \\
 & \href{https://github.com/RoaringBitmap/RoaringBitmap/commit/abdd6ff}{abdd6ff} & 1.14 & \cellcolor{lightred}{0.99} & 1.07 & 1.01 & 1.06 & \cellcolor{lightgreen}{1.17} & 1.06 & \cellcolor{lightred}{0.99} & 1.07 \\
 & \href{https://github.com/RoaringBitmap/RoaringBitmap/commit/1de6825}{1de6825} & 1.11 & 1.11 & 1.11 & 1.11 & 1.11 & - & 1.11 & - & - \\
 \midrule
 & Avg & 1.92 & 1.62 & 1.53 & 1.38 & 1.74 & 1.15 & 1.47 & 1.36 & 1.16 \\
\midrule
\midrule
 & Overall Avg & 5.86 & 1.94 & 2.62 & 2.51 & 3.42 & 1.18 & 1.45 & 1.67 & 1.72 \\
\bottomrule
\end{tabular}
}
\end{table}

Table~\ref{tab:hash_performance_summary} details the representative $pss$ scores achieved by the commercial model (Gemini 2.5 Pro) and the reasoning-enhanced open-weight model (DeepSeek R1) for individual tasks.  Note that 21 tasks where neither model was able to generate a plausible patch are omitted. Complete results for all models are available in our online appendix \footnote{\label{fn:rptable}\url{https://anonymous.4open.science/r/BenchLLMRealWorld/Analysis/RQ1/Complete_table3.pdf}}.

To highlight differences, cells are color-coded: a green background indicates cases where an LLM solution outperforms the developer solution, whereas a red background indicates a performance regression ($pss < 1.00$). White cells denote valid improvements that nonetheless fall short of the developer's solution. A value of ``-'' indicates that no plausible patch was generated in three attempts. Patches are grouped by project and sorted by the size of the developer solution's speedup. The average scores reported for each project and overall are calculated using the geometric mean~\cite{gm:04}, which is appropriate for aggregating normalized ratios.

Our overall finding that LLM-generated code often improves performance, but frequently to a lesser extent than what a human developer could achieve, is validated. For example, in task \texttt{bfb18d6} (\texttt{Presto}), the developer achieved a massive $pss$ of $145169.44$ (by far the largest improvement in our study), whereas most LLM solutions only achieve minor improvements. However, Gemini is able to achieve improvements close to the developer fix if provided both forms of context. The developer's strategy was to allow short-circuiting of an expensive intersection calculation, a bottleneck that DeepSeek R1 failed to identify, and Gemini only resolved with maximum guidance. A similar pattern of LLM patches improving performance, but less than the human developer, can be observed across numerous tasks, suggesting that LLMs can identify and implement effective optimizations, but often fail to find truly optimal solutions. A comparable situation can be observed for \texttt{ee74e4b} (\texttt{Presto}), where the human developer achieved a $pss$ of $38.54$ by identifying a major performance bottleneck, but the LLM solutions all performed around the baseline, only applying unrelated and ineffective micro-optimizations. 

There are also tasks where LLM solutions result in performance regressions. In particular, tasks in \texttt{Presto} commonly lead to regressions. One example is \texttt{9e010a7} (\texttt{Presto}), where all representative LLM solutions yielded lower performance. Interestingly, in this task, Gemini introduced a regression ($0.87$) caused by Gemini creating an extra method whose own overhead negated the optimization benefit, which is difficult to predict without runtime benchmarking. A more severe case of negative optimization is \texttt{5a9b28a} (\texttt{Netty}), where DeepSeek's solutions degraded performance to as low as $0.34$x the baseline, while the human developer achieved a $1.21$x speedup by simply switching queue polling from FIFO to LIFO to improve CPU cache locality (whereas DeepSeek incorrectly diagnosed a thread synchronization issue). Its aggressive refactoring of the hot \texttt{recycle} method likely bloated the bytecode and disabled JIT inlining, illustrating yet again how structural changes without profiling or benchmarking can destroy performance in highly-optimized code.

Table~\ref{tab:hash_performance_summary} also indicates that LLM performance differs significantly from project to project. In \texttt{Kafka}, we see many tasks where LLMs discover solutions that outperform developers. For instance, in task \texttt{aecd47b} (\texttt{Kafka}), the developer's patch achieved a minor $1.05$ speedup. In contrast, multiple solutions generated by both Gemini and DeepSeek R1 substantially surpassed the developer, with the LLM implementing a classic chunking pattern to reduce method call overhead while the developer used a more complex, platform-specific optimization. Another striking example is task \texttt{922a95a} (\texttt{Kafka}), where the developer's patch yielded a $1.28$ speedup, whereas Gemini, when guided with benchmark information, identified a novel optimization strategy that resulted in a substantial $8.81$ speedup. The developer reduced lock contention by splitting locks, but Gemini's strategy was to refactor the logic to avoid acquiring the lock altogether in certain cases. These examples suggest that LLMs can, in some cases, explore a different solution space than human developers, potentially uncovering non-obvious optimizations.

\begin{summary}
\textbf{Performance Improvement Evaluation (RQ1):} LLMs can improve the performance of real code, but statistically lag behind human developers. Commercial closed models generally outperforming open-weight alternatives. A meaningful description of the performance problem in the prompt is crucial for effective LLM optimization. The overall landscape is characterized by high volatility---models can degrade performance on some tasks, but can also occasionally discover novel strategies that surpass human experts.
\end{summary}

\subsection{Do LLMs Propose Similar Performance Improvements to Developers? (RQ2)}\label{sec:rq3_results}

We now discuss \emph{what kinds} of improvements LLMs suggest, and how similar they are to the developer baseline. To do so, we manually labeled each generated LLM solution using one of three codes (match, alignment, and divergence, following the taxonomy detailed in Section~\ref{sec:labeling}).

\subsubsection{Quantitative Comparison of Developer and LLM-Generated Patches}

Before categorizing optimization strategies, we first analyze the complexity of the generated patches compared to those written by developers. Table~\ref{tab:patch_complexity} presents the Lines of Code (LOC) modified, the number of continuous code blocks (Hunks) changes, and the change in Cyclomatic Complexity ($\Delta CC$)~\cite{mccabe1976complexity}, reporting both the range and the mean with standard deviation (SD). In terms of code volume, OpenAI o4-mini is notably concise ($19.9 \pm 15.9$). In contrast, Gemini and DeepSeek models exhibit high LOC changes with high standard deviations comparable to the developer baseline ($35.1 \pm 39.6$). This dispersion reflects the diverse nature of real-world optimization tasks and demonstrates that these models possess the capacity to generate extensive, context-heavy code (also recall that the \texttt{PerfOpt} tasks we used in this study may require changes in up to three separate files). In terms of the spatial distribution of changes, developers modify more disjoint code blocks ($3.7 \pm 3.2$ hunks) compared to LLMs ($2.1\text{--}3.0$), indicating that AI-generated patches remain more localized. Furthermore, regarding control flow, LLMs consistently introduce higher cyclomatic complexity ($\Delta CC$ mean $2.0\text{--}3.2$)  compared to developers ($1.4$), implying a tendency to optimize via elaborate logic rather than structural simplification.

\begin{table*}[h!]
\centering
\scriptsize
\caption{Complexity metrics of plausible patches.}
\label{tab:patch_complexity}
\begin{tabular}{l ccc ccc ccc}
\toprule
\multirow{2}{*}{\textbf{Source}} & \multicolumn{3}{c}{{LOC Modified}} & \multicolumn{3}{c}{{Hunks}} & \multicolumn{3}{c}{{$\Delta CC$}} \\
\cmidrule(lr){2-4} \cmidrule(lr){5-7} \cmidrule(lr){8-10}
 & {Min} & {Max} & {Mean $\pm$ SD} & {Min} & {Max} & {Mean $\pm$ SD} & {Min} & {Max} & {Mean $\pm$ SD} \\
\midrule
\rowcolor{gray!15} Developer & 2 & 220 & 35.1 $\pm$ 39.6 & 1 & 14 & 3.7 $\pm$ 3.2 & $-34$ & 26 & 1.4 $\pm$ 6.6 \\
\midrule
OpenAI o4-mini & 1 & 104 & 19.9 $\pm$ 15.9 & 1 & 15 & 2.1 $\pm$ 1.6 & $-10$ & 58 & 2.3 $\pm$ 7.1 \\
Gemini 2.5 Pro & 1 & 336 & 32.5 $\pm$ 32.6 & 1 & 14 & 2.9 $\pm$ 2.1 & $-26$ & 128 & 2.0 $\pm$ 8.3 \\
DeepSeek V3 & 2 & 255 & 36.8 $\pm$ 36.6 & 1 & 10 & 2.8 $\pm$ 2.1 & $-34$ & 44 & 3.0 $\pm$ 8.0 \\
DeepSeek R1 & 2 & 287 & 44.3 $\pm$ 43.5 & 1 & 9 & 3.0 $\pm$ 2.2 & $-26$ & 29 & 3.2 $\pm$ 6.6 \\
\bottomrule
\end{tabular}
\end{table*}

\subsubsection{Qualitative Comparison}

Figure~\ref{fig:rq3_result_snippet} presents a concrete example from our dataset. The performance task is to optimize a function that was inefficient for very small arrays. The developer's solution (Figure~\ref{fig:rq3_result_snippet} (a)) implements a fast path to reduce the overhead of creating a hash set. The ``strategy match'' (Figure~\ref{fig:rq3_result_snippet} (b)) mirrors this logic, implementing a functionally identical solution, despite differences in how they construct the final block. The ``strategy alignment'' (Figure~\ref{fig:rq3_result_snippet} (c)) identifies the same bottleneck, but employs a more general $O(n^2)$ nested loop for any array up to size 16. In another instance, the LLM produced a strategy divergence (Figure~\ref{fig:rq3_result_snippet} (d)), ignoring the small-size issue and instead implementing a different algorithm for arrays with no duplicates.

\begin{figure}[h!]
    \centering
    \begin{minipage}[t]{0.48\textwidth}
        \begin{lstlisting}[style=DiffStyle, title={{\scriptsize (a) Developer Solution}}]
--- a/path/to/ArrayDistinctFunction.java
+++ b/path/to/ArrayDistinctFunction.java
@@ -72,10 +72,19 @@ public final class ArrayDistinctFunction
     public static Block distinct(Type type, Block array)
     {
-        if (array.getPositionCount() == 0) {
+        if (array.getPositionCount() < 2) {
             return array;
         }
+        if (array.getPositionCount() == 2) {
+            if (type.equalTo(array, 0, array, 1)) {
+                return array.getSingleValueBlock(0);
+            }
+            else {
+                return array;
+            }
+        }
         TypedSet typedSet = new TypedSet(...);
         BlockBuilder blockBuilder = type.createBlockBuilder(...);
        \end{lstlisting}
    \end{minipage}
    \hfill
    \begin{minipage}[t]{0.48\textwidth}
        \begin{lstlisting}[style=DiffStyle, title={{\scriptsize (b) Strategy Match}}]
--- a/path/to/ArrayDistinctFunction.java
+++ b/path/to/ArrayDistinctFunction.java
@@ -72,13 +72,21 @@ public final class ArrayDistinctFunction
     public static Block distinct(Type type, Block array)
     {
-        if (array.getPositionCount() == 0) {
+        int positionCount = array.getPositionCount();
+        if (positionCount <= 1) {
             return array;
         }
+        if (positionCount == 2) {
+            if (type.equalTo(array, 0, array, 1)) {
+                return array.getSingleValueBlock(0);
+            }
+            return array;
+        }
         TypedSet typedSet = new TypedSet(...);
         BlockBuilder blockBuilder = type.createBlockBuilder(...);

        \end{lstlisting}
    \end{minipage}

    \begin{minipage}[t]{0.48\textwidth}
        \begin{lstlisting}[style=DiffStyle, title={{\scriptsize (c) Strategy Alignment}}]
--- a/path/to/ArrayDistinctFunction.java
+++ b/path/to/ArrayDistinctFunction.java
@@ -76,9 +76,28 @@ public final class ArrayDistinctFunction
-        TypedSet typedSet = new TypedSet(...);
-        for (int i = 0; i < array.getPositionCount(); i++) {
+        int count = array.getPositionCount();
+        // fast path for small arrays
+        if (count <= 16) {
+            for (int i = 0; i < count; i++) {
+                boolean found = false;
+                for (int j = 0; j < i; j++) {
+                    if (type.equalTo(array, i, array, j)) {
+                        found = true;
+                        break;
+                    }
+                }
+                if (!found) {
+                    type.appendTo(array, i, blockBuilder);
+                }
+            }
+            return blockBuilder.build();
+        }
+
+        // fallback to hash-based deduplication
+        TypedSet typedSet = new TypedSet(type, count);
+        for (int i = 0; i < count; i++) {
             if (!typedSet.contains(array, i)) {
        \end{lstlisting}
    \end{minipage}
    \hfill
    \begin{minipage}[t]{0.48\textwidth}
        \begin{lstlisting}[style=DiffStyle, title={{\scriptsize (d) Strategy Divergence}}]
--- a/path/to/ArrayDistinctFunction.java
+++ b/path/to/ArrayDistinctFunction.java
@@ -72,19 +72,27 @@ public final class ArrayDistinctFunction
     public static Block distinct(Type type, Block array)
     {
-        if (array.getPositionCount() == 0) {
+        if (array.getPositionCount() <= 1) {
             return array;
         }

         TypedSet typedSet = new TypedSet(...);
         BlockBuilder blockBuilder = type.createBlockBuilder();
+        boolean hasDuplicates = false;
         for (int i = 0; i < array.getPositionCount(); i++) {
-            if (!typedSet.contains(array, i)) {
+            if (typedSet.contains(array, i)) {
+                hasDuplicates = true;
+            }
+            else {
                 typedSet.add(array, i);
                 type.appendTo(array, i, blockBuilder);
             }
         }

+        if (!hasDuplicates) {
+            return array;
+        }
        \end{lstlisting}
    \end{minipage}

    \caption{A representative example of a strategy match, alignment, and divergence.}
    \label{fig:rq3_result_snippet}
\end{figure}

\begin{figure}[t!]
    \centering
    \includegraphics[width=0.7\linewidth]{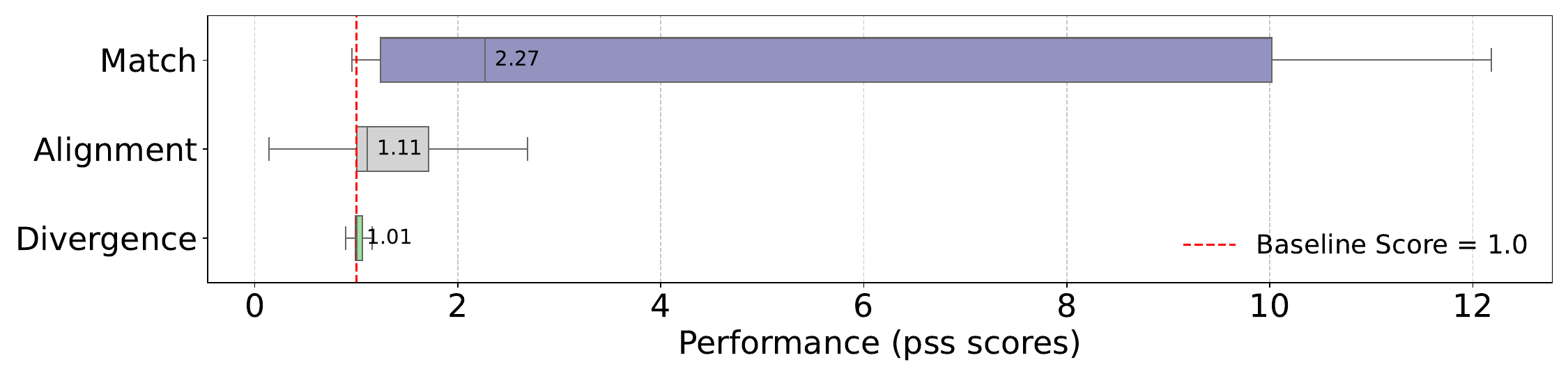}
    \caption{Distribution of $pss$ scores for each category, showing the performance impact of alignment. For visual clarity, outliers (identified using the IQR method~\cite{wan2014estimating}) are omitted from the boxplots.}
    \label{fig:boxplot_classification_analysis}
\end{figure}

Figure~\ref{fig:boxplot_classification_analysis} illustrates the performance impact based on the the novelty of the chosen optimization strategy. Strategy match solutions exhibit the highest median performance improvement ($2.27$). Strategy alignment solutions yield substantially lower improvement (median of $1.11$). Finally, strategy divergence solutions yield by far the lowest improvement, with a median of $1.01$. The variance in performance lowers as the degree of divergence increases, indicating that alternative approaches rarely result in substantial performance improvements. Rather, large-scale improvements generally require developers and LLMs to identify the same optimization opportunity. This is to be expected, as real-life production code is unlikely to have multiple differing avenues for substantial optimization.

\begin{figure}[h!]
    \centering
    \includegraphics[width=1\linewidth]{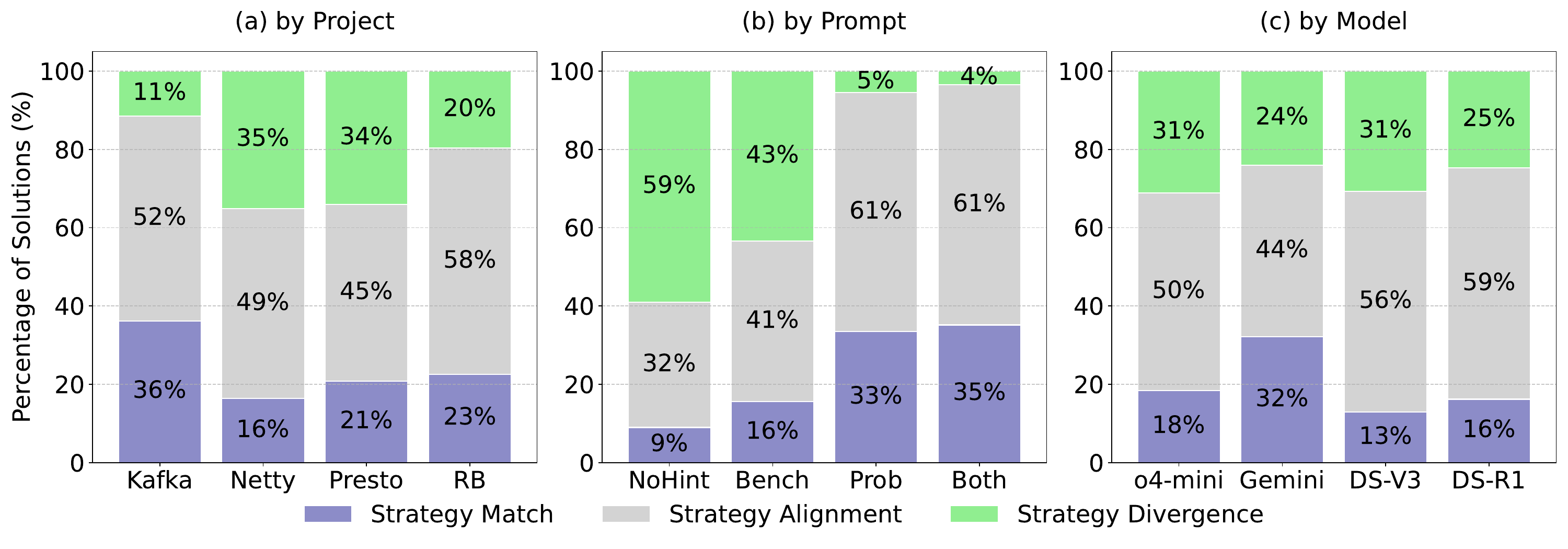}
    \caption{Categorization of the similarity of LLM-generated solutions to developer solutions, broken down by (a) project, (b) prompting strategy, and (c) model.}
    \label{fig:classification_analysis}
\end{figure}

Figure \ref{fig:classification_analysis} depicts the results of our categorization by (a) project, (b) prompt, and (c) model. All three aspects have an effect on the distribution of solutions. 
First, strategy alignment was the most prevalent category across all four projects, ranging from 45\% in \texttt{Presto} to 58\% in \texttt{RoaringBitmap}. However, clear distinctions remain: \texttt{Kafka} exhibited the highest proportion of strategy matches (36\%), whereas \texttt{Netty} and \texttt{Presto} showed the highest rates of strategy divergence (35\% and 34\%, respectively). This implies that strict strategy alignment becomes increasingly difficult as project complexity rises. In complex code bases like \texttt{Netty}, models struggle to infer the developer's optimization intent, instead defaulting to generic or hallucinated optimization patterns. 

We observe distinct differences in LLM-developer similarity based on the prompting strategy (Figure~\ref{fig:classification_analysis}(b)). The transition across strategies reveals that explicit problem descriptions serve primarily as a constraint on the solution space. When diagnostic information is provided (``Problem'' and ``Both''), the rate of strategy divergence collapses from 59\% (in ``No Hint'') to a negligible 5\%. However, this guidance does not necessarily enable the model to replicate the expert's optimal solution. Instead, the distribution shifts heavily toward \textit{strategy alignment} (reaching 61\% for both strategies) rather than \textit{strategy match} (which plateaus at 33--35\%). This indicates that while knowing the root cause effectively prevents models from addressing unrelated code (divergence), it typically steers them toward a valid but sub-optimal algorithmic fix (alignment) rather than the precise optimization implemented by developers. Conversely, providing benchmarks alone (``Bench'') proves insufficient to act as a similar guardrail, resulting in a persistent 43\% divergence rate. This reinforces a critical limitation---without explicit semantic cues to locate the bottleneck, models struggle to autonomously profile code, often mistaking intended logic for performance defects.

Finally, we also observed (Figure~\ref{fig:classification_analysis}(c)) that Gemini produced the highest proportion of strategy matches (32\%), suggesting its approach is more frequently identical to that of a human developer. In contrast, the open-weight DeepSeek models (DS-V3 and DS-R1) exhibited the lowest rates of strategy matches (13\% and 16\%, respectively). Instead, these models predominantly fell into the strategy alignment category (56\% and 59\%), implying that while they are effective at finding the core issue, their fixes consistently differ from the human baseline. Comparing DS-V3 and DS-R1 reveals that reasoning capabilities reduced divergence (31\% to 25\%) but did not increase matches. This indicates that explicit reasoning helped the model stay on topic, but provided no clear advantage in identifying optimal fixes.

\begin{summary}
\textbf{Optimization Strategy (RQ2):} LLM-generated patches are structurally more complex and localized than developer solutions, yet the most effective optimizations occur when models closely mimic human strategies. Divergence from the developer's approach correlates with diminishing returns, highlighting a dual deficiency in the current generation of models: (1) a lack of intrinsic profiling capabilities to locate bottlenecks, and (2) a limited capacity to synthesize optimal improvements even when the target context is provided (e.g., because a human or upstream agent benchmarked and diagnosed the issue upfront). 
\end{summary}

\section{Discussion}

Our findings offer key implications for the potential role of LLMs in performance engineering. In this section, we discuss the practical applications for developers, the need for integrated tooling, and suggestions for future research. We also examine the limitations of our study.

\subsection{Implications}

\smallskip\noindent\textbf{Benchmarking Beyond Algorithms:}
Our study highlights a significant disconnect between current evaluation standards and real-world problems---real performance issues are fundamentally more diverse than those typically found in academic or competitive programming datasets. Recent benchmarks often focus on isolated, algorithmic inefficiencies~\cite{niu:24,peng:25}, and they also report that LLMs struggle as context grows---reporting a sharp drop in GPT-4o's average speed-up from $8.28\times$ (function-level) to $1.11\times$ (file-level). It is crucial to note that their file-level tasks still rely primarily on data from algorithmic competitions (e.g., APPS, Code Contests). However, our analysis of \texttt{PerfOpt} reveals that real-world performance bottlenecks arise not just from the algorithmic structure of the code, but from system-specific problems including concurrency issues, suboptimal API usage, or memory allocation patterns. As exemplified by a case (\texttt{Netty} 5a9b28a) in Section~\ref{Task-Specific}, this complexity often manifests non-linearly---rather than requiring a full algorithmic refactoring, a drastic performance shift can be achieved by identifying and rewriting 1--2 lines of code. 

However, pinpointing such precise hotspots requires deep diagnostic reasoning rather than generic algorithmic design skills. Despite employing state-of-the-art commercial models (e.g., Gemini 2.5 Pro), we found that without explicit guidance, models struggled significantly. Further, unlike the consistent gains typically reported in algorithmic tasks, our analysis revealed extreme volatility, where performance ranged from discovering novel strategies to causing severe regressions. This disparity suggests that solely evaluating LLMs on algorithmic challenges yields an overly optimistic assessment of their readiness. To meaningfully advance the field, future research must prioritize the use of realistic, validated datasets that capture the multifaceted performance problems encountered by developers in the field. \texttt{PerfOpt} can serve as a starting point for such a dataset.

\smallskip\noindent\textbf{The Role of LLMs in Performance Engineering:} Our findings indicate that while current LLMs possess substantial code generation capabilities, they lack intrinsic intuition or runtime profiling capabilities required for autonomous optimization. Without a clear context for the underlying performance issue, LLM solutions tend to be shortsighted, sometimes applying micro-optimizations that miss the core issue. This suggests that their ``creativity'' may be more akin to a shallow pattern-matching of alternative code structures rather than a deep understanding of performance bottlenecks. While LLMs yielded better solutions when offered problem context as part of the prompt, this gap in deep performance reasoning is a critical area for future research. 

Despite these limitations, we observed rare-but-significant instances where an LLM proposed an optimization that an experienced developer missed. This highlights the potential for LLMs as a tool for brainstorming alternative optimization strategies. Therefore, the most immediate and practical role for LLMs in this domain is as a performance assistant that suggests potential optimizations, which a human developer must then critically evaluate, benchmark, and validate before integration. However, we also see potential in devising agent-based performance optimization systems, which integrate separate LLM-based tools for runtime profiling, benchmarking, and diagnostics, to overcome the limitations our study has revealed.

\subsection{Threats to Validity}

\noindent\textbf{External Validity:} Our finding are constrained by our choice of models and dataset. We mitigate this risk by selecting models representing multiple sizes and licensing models (i.e., commercial and open-weight)\footnote{Regarding model deployment, although DeepSeek is an open-weight model capable of local execution, we accessed it via its official API. However, our results should be replicable if one downloads and locally executes the model.} and by selecting representative real-world projects with developer-written patches and benchmarks. Still, our results may vary for other languages or domains with different optimization patterns.

\smallskip\noindent\textbf{Internal Validity:} The primary internal threat is the reliability of performance measurements in a cloud environment. We mitigated transient noise by executing benchmarks in isolation using the JMH framework, which is designed to minimize JVM warmup effects, and by reporting the mean of multiple trials. Another potential concern is data leakage, given that the open-source projects used may appear in the training data of the models. However, our results in RQ2 provide strong evidence against memorization being the primary driver of performance. The produced patches differ substantially in their structure (see Table~\ref{tab:patch_complexity}). Further, in the ``No Hint'' setting, models failed to replicate the developer's strategy in most cases. If the models were simply recalling the ground-truth fix from their training data, we would expect a higher rate of strategy matching---even without explicit context---and structural metrics more similar to the developer patches. 

\smallskip\noindent\textbf{Construct Validity:} We assess patch plausibility by ensuring a patch passes the provided unit test suite, rather than guaranteeing full semantic correctness. This is a common practice, but it is limited by the completeness of the test suite itself, which may not cover all edge cases. Further, when using statistical analyses, we have attempted to ensure the base assumptions behind these analyses are met. We have favored nonparametric methods, as distribution characteristics are not generally known a priori, and normality cannot be assumed.

\section{Conclusions}

In this paper, we bridge the gap between synthetic algorithmic benchmarks and real-world performance engineering by introducing \texttt{PerfOpt}. We presented an empirical study to assess the capability of modern LLMs to optimize software performance. Our evaluation reveals that---while modern LLMs possess substantial code generation capabilities, at times outperforming human developers---they also fundamentally struggle with the diagnostic reasoning required for real-world systems. Unlike the consistent speedups reported in purely-algorithmic contexts, our results highlight a landscape of extreme volatility, where unguided models frequently misdiagnose bottlenecks and diverge from effective strategies. This demonstrates that current SOTA models act as high-variance search engines rather than reliable engineers. Consequently, their optimal role is not autonomous replacement but collaborative acceleration, where human experts provide the diagnosis, benchmark, and potential pathways to a solution, and LLMs synthesize the implementation (i.e., a situation similar to the ``Both'' prompt).

Alternatively, our results motivate the development of a multi-agent system~\cite{idrisov:25}, where AI agents apply tools and perform additional analyses as input to patch generation, including profiling and runtime benchmarking of proposed improvements. Additionally, we argue that the field requires the expansion of realistic benchmarks like \texttt{PerfOpt} to encompass broader language ecosystems and architectural patterns, and move empirical LLM performance research beyond artificial code examples.

\section*{Acknowledgment}

Grant acknowledgments removed for double-blind review. 
We additionally acknowledge the use of Generative AI to improve the grammar of the manuscript. 

\section*{Data Availability}
Our dataset, measurement results, and raw data are available in our online appendix.\footnote{\label{fn:rp}\url{https://anonymous.4open.science/r/BenchLLMRealWorld/}}

\bibliographystyle{ACM-Reference-Format}
\bibliography{bibliography.bib}

\end{document}